\documentclass[twocolumn]{aastex62}
\usepackage{natbib}
\usepackage{epsfig}
\usepackage{graphicx}
\usepackage{subfigure}
\usepackage{float}
\usepackage{amsmath}
\usepackage{color}
\usepackage{amssymb}
\usepackage{amsfonts}
\usepackage{apjfonts}
\newcommand{\PMO}{Purple Mountain Observatory, Chinese Academy of Sciences, Nanjing 210023, China; jjwei@pmo.ac.cn, xfwu@pmo.ac.cn}
\newcommand{\USTC}{School of Astronomy and Space Science, University of Science and Technology of China, Hefei 230026, China}
\newcommand{\GXU}{Guangxi Key Laboratory for Relativistic Astrophysics, Nanning 530004, China}
\newcommand{\LDM}{Key Laboratory of Dark Matter and Space Astronomy, Purple Mountain Observatory, Chinese Academy of Sciences Nanjing 210034, China; zhd@pmo.ac.cn}
\newcommand{\PKU}{Department of Astronomy, School of Physics, Peking University, Beijing 100871, China}

\shortauthors{Lan et al.}

\begin{document}

\title{The Stellar-mass Function of Long Gamma-Ray Burst Host Galaxies}


\author{Guang-Xuan Lan}
\affiliation{\PMO}
\affiliation{\USTC}
\affiliation{\GXU}
\affiliation{\PKU}

\author{Jun-Jie Wei}
\affiliation{\PMO}
\affiliation{\USTC}
\affiliation{\GXU}

\author{Ye Li}
\affiliation{\PMO}

\author{Hou-Dun Zeng}
\affiliation{\PMO}
\affiliation{\LDM}

\author{Xue-Feng Wu}
\affiliation{\PMO}
\affiliation{\USTC}

\begin{abstract}
Long gamma-ray bursts (GRBs) have been discussed as a potential tool to probe the cosmic star formation rate (SFR)
for a long time. Some studies found an enhancement in the GRB rate relative to the galaxy-inferred SFR
at high redshifts, which indicates that GRBs may not be good tracers of star formation. However, in these
studies, the GRB rate measured at any redshift is an average over all galaxies at that epoch. A deep
understanding of the connection between GRB production and environment also needs to characterize the population
of GRB host galaxies directly. Based on a complete sample of GRB hosts, we constrain the stellar-mass function (SMF)
of GRB hosts, and examine redshift evolution in the GRB host population. Our results confirm that a strong redshift
evolution in energy (with an evolution index of $\delta=2.47^{+0.73}_{-0.89}$) or in density ($\delta=1.82^{+0.22}_{-0.59}$)
is needed in order to account for the observations. The GRB host SMF can be well described by the
Schechter function with a power-law index $\xi\approx-1.10$ and a break mass $M_{b,0}\approx4.9\times10^{10}$ ${\rm M}_\odot$,
independent of the assumed evolutionary effects. This is the first formulation of the GRB host SMF.
The observed discrepancy between the GRB rate and the galaxy-inferred SFR may also be explained by an evolving SMF.
\end{abstract}


\section{Introduction}
As the extremely powerful explosions of massive stars \citep{2006ARA&A..44..507W}, long-duration gamma-ray bursts (GRBs) are
detectable out to very high redshifts \citep{2009Natur.461.1258S,2009Natur.461.1254T,2011ApJ...736....7C}. The redshift
distribution of long GRBs and the properties of their host galaxies can therefore be used to probe the evolution of the
star formation rate (SFR) density with cosmic time (e.g., \citealt{1997ApJ...486L..71T,1998MNRAS.294L..13W,2000ApJ...536....1L,
2001ApJ...548..522P}). Relevant studies in this field have been carried out for more than two decades, but it remains
an open question whether long GRBs can be good tracers of cosmic star formation. By studying the GRB redshift distribution,
there is a general agreement about the fact that the comoving GRB rate does not purely follow the cosmic SFR,
but exhibits an enhancement of GRB detections at high redshifts ($z\ga2-3$) with respect to the expectation from
the star formation history (e.g., \citealt{2006ApJ...647..773D,
2007JCAP...07..003G,2007ApJ...661..394L,2008ApJ...673L.119K,2008MNRAS.388.1487L,2010MNRAS.406..558Q,2010MNRAS.406.1944W,
2011MNRAS.417.3025V,2012ApJ...745..168L,2012ApJ...744...95R,2013A&A...556A..90W,2014ApJ...783...24L,2014MNRAS.439.3329W,
2015ApJ...801..102P,2016ApJ...817....8P,2015A&A...581A.102V,2016A&A...590A.129J,2017IJMPD..2630002W,2019A&A...623A..26P,
2019MNRAS.488.4607L,2021MNRAS.508...52L,2022arXiv220606390G}).
Several possible mechanisms have been proposed to reconcile the discrepancy between the GRB rate and the SFR, such as
cosmic metallicity evolution \citep{2006ApJ...638L..63L,2008MNRAS.388.1487L,2012ApJ...749...68S,2019MNRAS.488.5823L,2022arXiv220606390G},
an evolving initial mass function of stars \citep{2009ChA&A..33..151X,2011ApJ...727L..34W}, and an evolving luminosity
function of GRBs (e.g., \citealt{2002ApJ...574..554L,2004ApJ...611.1033F,2007ApJ...656L..49S,2009MNRAS.396..299S,
2012ApJ...749...68S,2011MNRAS.416.2174C,2015MNRAS.454.1785T,2016A&A...587A..40P,2019MNRAS.488.4607L,2021MNRAS.508...52L}).

Among the possible mechanisms, cosmic metallicity evolution appears to be a compelling one. The progenitors of long GRBs
are expected to be preferentially localized in low-metallicity environments, where the loss of mass and angular momentum are
lessened, allowing a relativistic jet to form \citep{1999ApJ...524..262M,2005A&A...443..581H,2005A&A...443..643Y,
2006ApJ...637..914W,2006A&A...460..199Y}.
And indeed it has been shown that long GRBs tend to occur in lower metallicity (as well as lower stellar-mass) galaxies,
with metallicity $Z\leq0.3Z_{\odot}$ (see \citealt{2016SSRv..202..111P} and references therein). However, some works
suggested that GRBs could occur in high-metallicity host environments (e.g.,
\citealt{2010ApJ...712L..26L,2012A&A...539A.113E,2013A&A...556A..23E,2012MNRAS.420..627S,2013ApJ...772...42H}).
It is important to note that previous studies of the GRB redshift distribution are necessarily imprecise.
Specifically, their derived GRB rate at any redshift is an average over all galaxies at that epoch (a diverse population
covering orders of magnitude in metallicities, SFRs, etc.) and cannot disclose which types of galaxies contribute
most to the GRB rate. To thoroughly understand the link between GRB production and environment, one also is required to
characterize the population of GRB host galaxies directly \citep{2016ApJ...817....8P}.

One of the difficulties in attempting to connect GRB observations and their environments to particular progenitor
systems is obtaining a complete sample of GRB host galaxies. Recently, \cite{2016ApJ...817....7P,2016ApJ...817....8P}
carried out a new survey of the GRB host galaxy population. Their survey provides the first host galaxy sample that
is unbiased and sufficiently large to statistically examine redshift evolution in the host galaxy population in detail.
In this work, we make use of the complete sample presented in \cite{2016ApJ...817....7P,2016ApJ...817....8P} to
quantify the stellar-mass function (SMF) of GRB host galaxies. Meanwhile, we use the observed redshift distribution
of the complete sample to probe and constrain the evolution of the GRB host population in redshift.

The structure of the paper is as follows: In Section~\ref{sec:sample}, we introduce the GRB host sample used for our analysis.
We describe our analysis method in Section~\ref{sec:method}, while the model results are presented in Section~\ref{sec:result}.
Finally, in Section~\ref{sec:summary} we draw our conclusions. Throughout this work we adopt a flat $\Lambda$CDM
model with cosmological parameters $H_{0}=70$ km s$^{-1}$ Mpc$^{-1}$, $\Omega_{\rm m}=0.3$, and $\Omega_{\Lambda}=0.7$.

\section{The Sample}
\label{sec:sample}
The redshifts of about 40\% of all GRBs detected by the {\it Swift} satellite have been measured \citep{2021MNRAS.508...52L}.
Though this is a significant improvement compared to the pre-{\it Swift} era, from the point of view of characterizing the GRB
host population or redshift distribution, the sample is still far from being considered unbiased and complete.
The {\it Swift} GRB Host Galaxy Legacy Survey (SHOALS) built a new unbiased sample of long GRB host galaxies,
which is composed of 119 GRBs at $0.03<z<6.29$ drawn from the {\it Swift} GRB catalog using a series of selection criteria and
requiring the {\it Swift}/Burst Alert Telescope (BAT) fluence to exceed $S>10^{-6}$ erg $\rm cm^{-2}$ \citep{2016ApJ...817....7P}.
Their optimized observability cuts
combined with host galaxy follow-up enabled redshift measurements for 110 of these bursts (representing a completeness of 92\%).
Subsequently, \cite{2016ApJ...817....8P} presented rest-frame near-infrared (NIR) luminosities and stellar masses for
this uniform sample of 119 GRB host galaxies through the {\it Spitzer} Infrared Array Camera (IRAC) observations.
As the first large, highly complete, and multiband survey of an unbiased GRB host galaxy sample, SHOALS provides
unprecedented insight into galaxy evolution and cosmic history \citep{2016ApJ...817....7P,2016ApJ...817....8P}.

\begin{figure*}
\center
\includegraphics[angle=0,scale=0.55]{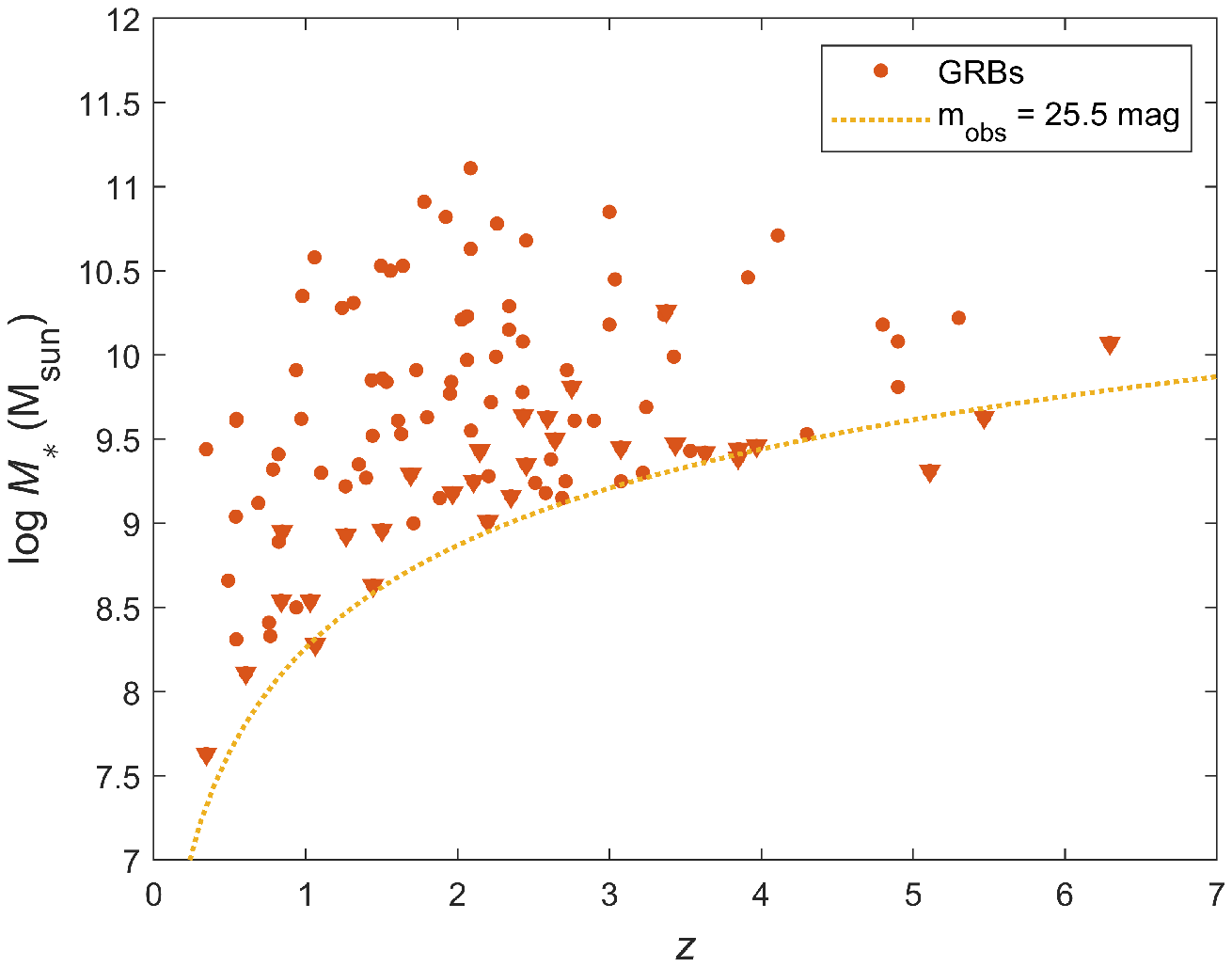}
\includegraphics[angle=0,scale=0.55]{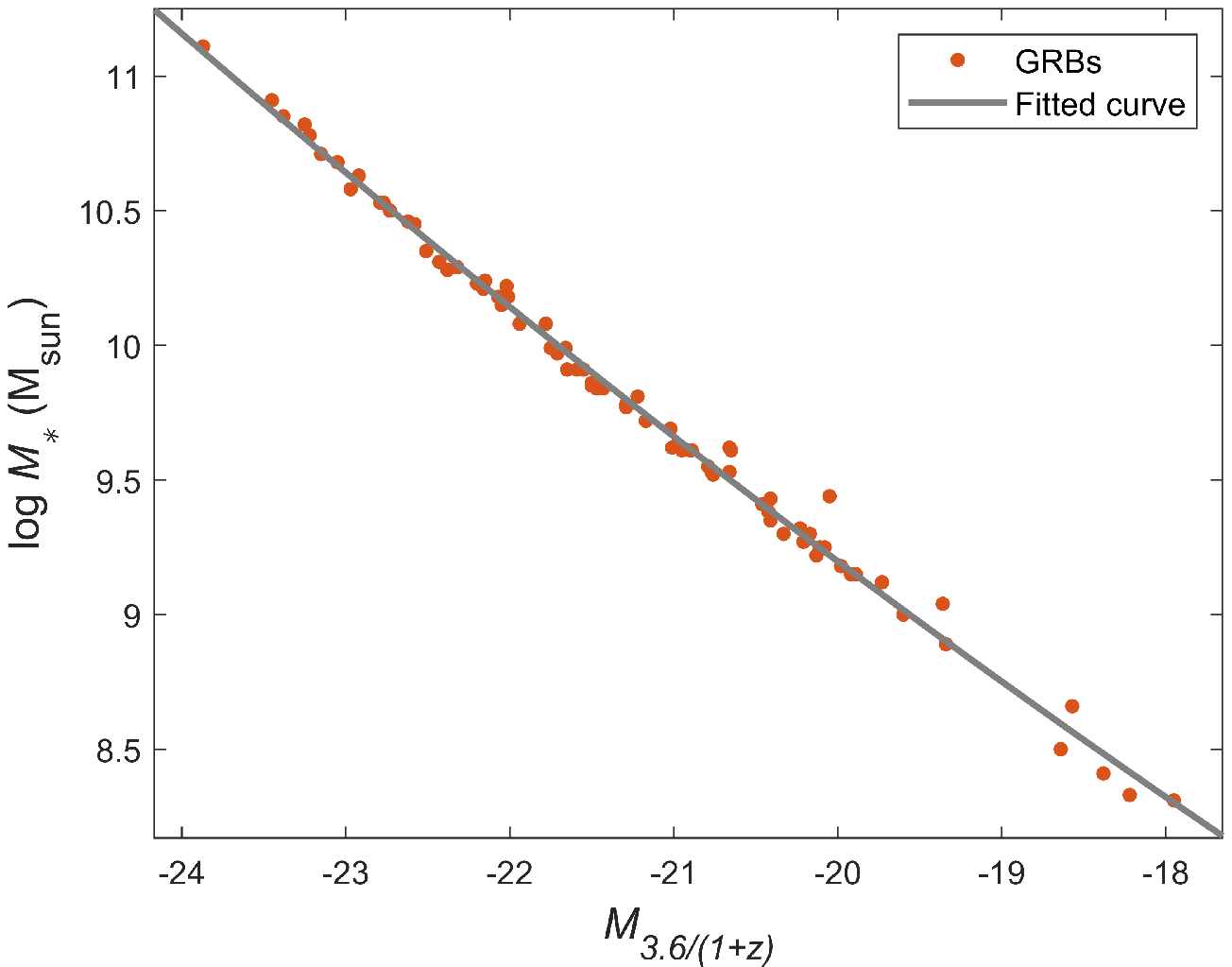} 
\caption{Left panel: the stellar mass--redshift distribution of 109 GRB host galaxies in the SHOALS sample. The inverted
triangles represent the hosts with stellar-mass upper limits. The dotted line represents the stellar-mass threshold adopted
in our calculations (corresponding to an apparent-magnitude limit $m_{\rm obs}=25.5$ mag; see text and Equations~(\ref{eq:MAB})
and (\ref{eq:M-MAB})). Right panel: stellar mass as a function of the absolute magnitude at a wavelength of
$3.6/(1+z)$ $\mu$m for 79 GRB host galaxies with measured stellar mass and redshift. The best-fitting curve is parameterized
in Equation~(\ref{eq:M-MAB}).}\label{fig1}
\end{figure*}

In this work, we use the SHOALS sample to investigate the stellar-mass distribution of GRB hosts across
cosmic history and its possible implications for interpreting the observed GRB redshift distribution.
We calculate the isotropic-equivalent energies $E_{\rm iso}$ in the 1--$10^{4}$ keV rest-frame energy range
for these 110 GRBs having redshifts. Note that there is a low-luminosity/energy burst (GRB 060218) with
$E_{\rm iso}\leq10^{49}$ erg. As the low-luminosity/energy bursts have been claimed to be a distinct population
\citep{2004Natur.430..648S,2006ApJ...645L.113C,2007MNRAS.382L..21C,2007ApJ...662.1111L}, GRB 060218 is discarded
in our analysis. The remaining sample includes 109 GRBs, whose stellar mass--redshift distribution is presented
in the left panel of Figure~\ref{fig1}. It is also noteworthy that 30 out of these 109 GRB host galaxies only have
stellar-mass upper limits \citep{2016ApJ...817....8P}. In principle, the galaxies with stellar-mass upper limits
should be removed in the determination of the stellar-mass distribution of GRB hosts. However, the direct abandonment of
these hosts would introduce a new bias on shaping the GRB redshift distribution. In the following, we will simply treat
the upper mass limits as the stellar-mass measurements for these 30 GRB hosts.

\section{Analysis Method}
\label{sec:method}

In order to infer the model-free parameters, a maximum likelihood method firstly proposed by \cite{1983ApJ...269...35M}
is adopted. The likelihood function $\mathcal{L}$ is defined as (e.g.,
\citealt{1998ApJ...496..752C,2006ApJ...643...81N,2009ApJ...699..603A,2012ApJ...751..108A,2010ApJ...720..435A,
2014MNRAS.441.1760Z, 2016MNRAS.462.3094Z,2019MNRAS.488.4607L,2021MNRAS.508...52L})
\begin{equation}
\mathcal{L}=\exp\left(-N_{\rm exp}\right)\prod ^{N_{\rm obs}}_{i=1} \Phi(E_{{\rm iso},i},\;M_{\ast,i},\;z_i,\;t_i)\;,
\label{eq:Likelihood}
\end{equation}
where $N_{\rm obs}$ is the number of the observed sample, $N_{\rm exp}$ is the expected number of GRB detections,
and $\Phi(E_{\rm iso},\;M_{\ast},\;z,\;t)$ is the observed rate of GRBs per unit time at redshift $z\sim z+\mathrm{d}z$
with energy $E_{\rm iso}\sim E_{\rm iso}+\mathrm{d}E_{\rm iso}$ and stellar mass $M_{\ast}\sim  M_{\ast}+\mathrm{d}M_{\ast}$,
which can be expressed as
\begin{eqnarray}
\Phi\left(E_{\rm iso},\;M_{\ast},\;z,\;t\right)&=& \nonumber\frac{\mathrm{d}^4 N}{\mathrm{d}t\mathrm{d}z\mathrm{d}M_{\ast}\mathrm{d}E_{\rm iso}}
=\frac{\mathrm{d}^4 N}{\mathrm{d}t\mathrm{d}V\mathrm{d}M_{\ast}\mathrm{d}E_{\rm iso}}\times\frac{\mathrm{d}V}{\mathrm{d}z}\\
&=& \frac{\Delta \Omega}{4\pi}\frac{\psi(z)}{1+z}\varphi(M_{\ast},z)\phi(E_{\rm iso})\times\frac{\mathrm{d}V}{\mathrm{d}z}\;,
\end{eqnarray}
where $\Delta \Omega=1.33$ sr is the $Swift$/BAT field of view, $\psi(z)$ is the comoving formation rate of GRBs
(in units of $\rm Mpc^{-3}$ $\rm yr^{-1}$), $(1+z)^{-1}$ accounts for the cosmological time dilation, $\phi(E_{\rm iso})$
is the normalized isotropic-equivalent energy function of GRBs, and $\varphi(M_{\ast},z)$ is the normalized SMF of
GRB hosts, which evolves with redshift dependents on the assumed model (see more below).
${\rm d}V(z)/{\rm d}z=4\pi c D_L^2(z)/[H_{0}(1+z)^2\sqrt{\Omega_{\rm m}(1+z)^3+\Omega_{\Lambda}}]$
is the comoving volume element in a flat $\Lambda$CDM model, where $D_L(z)$ is the luminosity distance.

Since in the collapsar model the births of GRBs just mean the deaths of short-lived massive stars, the GRB formation rate
$\psi(z)$ should in principle be related to the cosmic SFR $\psi_\star(z)$, i.e.,
\begin{equation}
\psi(z)=\eta \psi_\star(z)\;,
\end{equation}
where $\eta$ is the GRB formation efficiency in units of ${\rm M}_{\odot}^{-1}$. The SFR (in units of
${\rm M}_{\odot}$ $\rm yr^{-1}$ $\rm Mpc^{-3}$) can be parameterized as \citep{2006ApJ...651..142H,2008MNRAS.388.1487L}
\begin{equation}
\psi_{\star}(z)=\frac{0.0157+0.118z}{1+(z/3.23)^{4.66}}\;.
\end{equation}

For the GRB energy function $\phi(E_{\rm iso})$, we employ a broken power-law form
\begin{equation}
 \phi (E_{\rm iso})  = \frac{A}{\ln(10)E_{\rm iso}}\left\{\begin{array}{l}
{\left(\frac{{{E_{\rm iso}}}}{{{E_b}}}\right)^a};\,\,{E_{\rm iso}} \le {E_b}\\
{\left(\frac{{{E_{\rm iso}}}}{{{E_b}}}\right)^b};\,\,{E_{\rm iso}} > {E_b}\;,
\end{array} \right.
\end{equation}
where $A$ is a normalization factor, and $a$ and $b$ are the power-law indices before and after the break energy $E_b$.

Empirically, the galaxy SMF can be well described by a Schechter form \citep{2008ApJ...680...41D}.
Therefore, we use the same Schechter expression for the stellar-mass distribution of GRB host galaxies:
\begin{equation}
\varphi(M_{\ast})= \varphi_{0}\left(\frac{M_{\ast}}{M_{b,0}}\right)^{\xi} \exp \left(-\frac{M_{\ast}}{M_{b,0}}\right) \frac{1}{M_{b,0}}\;,
\label{eq:MF}
\end{equation}
where $\varphi_{0}$ is a normalization factor, $\xi$ is the power-law index, and $M_{b,0}$ is the break mass.

Considering the fluence threshold that is used to define the SHOALS sample (i.e., $S_{\rm lim}=10^{-6}$ erg $\rm cm^{-2}$
in the 15--150 keV), the expected number of GRB detections should be
\begin{equation}
\begin{aligned}
N_{\rm exp} = & \frac{\Delta \Omega T}{4\pi} \int ^{z_{\rm max}} _{0}  \frac{\psi(z)}{1+z} \frac{\mathrm{d}V(z)}{\mathrm{d}z} \mathrm{d}z\int ^{M_{\ast, \rm max}}_{M_{\ast, \rm lim}(z)}\varphi(M_{\ast}, z)\mathrm{d}M_{\ast}\\
&\times\int ^{E_{\rm max}}_{{\rm max}[E_{\rm min},E_{\rm lim}(z)]}\phi(E_{\rm iso})\mathrm{d}E_{\rm iso} \;,
\label{eq:Nexp}
\end{aligned}
\end{equation}
where $T\sim7$ yr is the observational period of {\it Swift} that covers the SHOALS sample and $M_{\ast,\rm lim}(z)$
represents the stellar-mass threshold of {\it Spitzer}-IRAC at redshift $z$. Since $z<10$ and
$M_{\ast}<10^{12}$ ${\rm M}_{\odot}$ for the current GRB host population, we adopt a maximum redshift $z_{\rm max}=10$
and a maximum stellar mass $M_{\ast,\rm max}=10^{12}$ ${\rm M}_{\odot}$. The energy function is assumed to extend
between minimum and maximum energies $E_{\rm min}=10^{51}$ erg and $E_{\rm max}=10^{56}$ erg. The energy threshold
appearing in Equation~(\ref{eq:Nexp}) can be computed by
\begin{equation}
E_{\rm lim}(z)=\frac{4\pi D_L^2(z)S_{\rm lim}}{1+z}  \frac{\int^{10^4/(1+z)\;{\rm keV}}_{1/(1+z)\;{\rm keV}} EN(E)\mathrm{d}E}{\int^{150\;{\rm keV}}_{15\;{\rm keV}} EN(E)\mathrm{d}E}\;,
\end{equation}
where $N(E)$ is the GRB photon spectrum. To describe the burst spectrum, we employ a typical Band function with low- and
high-energy spectral indices $-1$ and $-2.3$, respectively \citep{1993ApJ...413..281B,2000ApJS..126...19P,2006ApJS..166..298K}.
For a given isotropic-equivalent energy $E_{\rm iso}$, the peak energy of the spectrum $E_{p}$ is estimated through
the empirical $E_{p}$--$E_{\rm iso}$ correlation \citep{2002A&A...390...81A,2012MNRAS.421.1256N}, i.e.,
$\log \left[E_{p}(1+z)\right]=-29.60+0.61\log E_{\rm iso}$.

For each GRB host galaxy (with measured redshift) from the SHOALS sample, the absolute magnitude at a wavelength of
$3.6/(1+z)$ $\mu$m can be calculated by
\begin{equation}
M_{3.6/(1+z)}=m_{\rm obs}-\mu(z)+2.5\log(1+z)\;,
\label{eq:MAB}
\end{equation}
where $m_{\rm obs}$ is the observed IRAC aperture magnitude and $\mu(z)$ is the distance modulus at $z$.
It has been suggested that the quantity $M_{3.6/(1+z)}$ alone can be used as a reasonable stellar-mass
proxy \citep{2016ApJ...817....8P}. As shown in the right panel of Figure~\ref{fig1}, there is a strong
correlation between the absolute magnitudes $M_{3.6/(1+z)}$ and the logarithmic stellar masses $\log M_{\ast}$
of GRB host galaxies. For simplicity, here we use a second-order polynomial to model the $\log M_{\ast}$--$M_{3.6/(1+z)}$ relation
\begin{equation}
\log M_{\ast}=p_{0}+p_{1} M_{3.6/(1+z)}+p_{2} M_{3.6/(1+z)}^{2}\;,
\label{eq:M-MAB}
\end{equation}
where the best-fitting parameters are $p_0=3.59_{-1.13}^{+1.13}$ , $p_1=-0.11_{-0.11}^{+0.11}$, and
$p_2=0.0087_{-0.0025}^{+0.0025}$ at 68\% confidence level. The observed magnitude $m_{\rm obs}$ can then be
translated to stellar-masses $M_{\ast}$ at different $z$ via Equations~(\ref{eq:MAB}) and (\ref{eq:M-MAB}).
We show that the stellar-mass threshold $M_{\ast,\rm lim}$ appearing in Equation~(\ref{eq:Nexp}) may be approximated
well using a magnitude limit of {\it Spitzer}-IRAC, $m_{\rm obs, lim}=25.5$ mag, according to Equations~(\ref{eq:MAB})
and (\ref{eq:M-MAB}) (see the dotted line in the left panel of Figure~\ref{fig1}).

\begin{figure}
\center
\includegraphics[angle=0,scale=0.5]{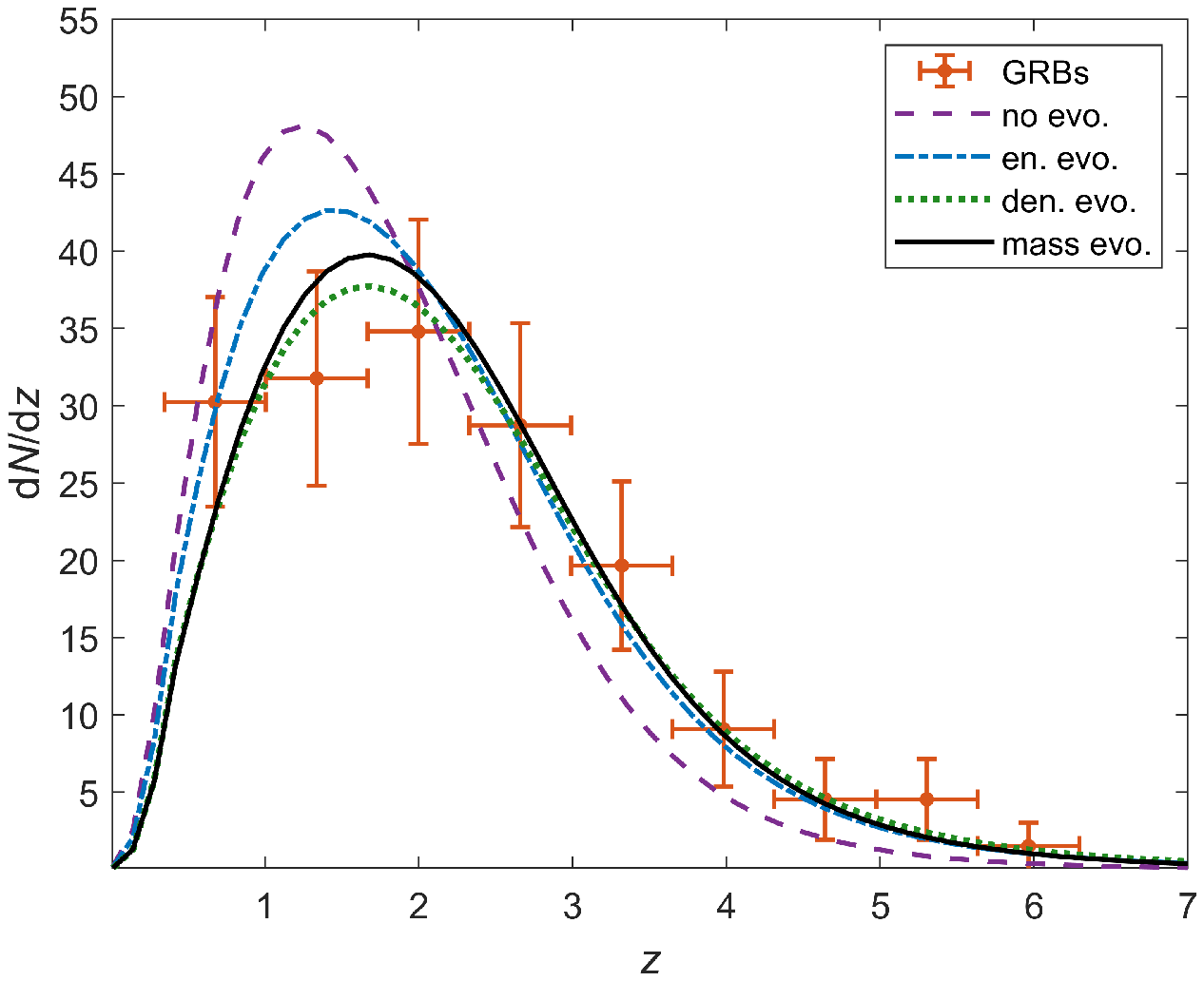} 
\includegraphics[angle=0,scale=0.5]{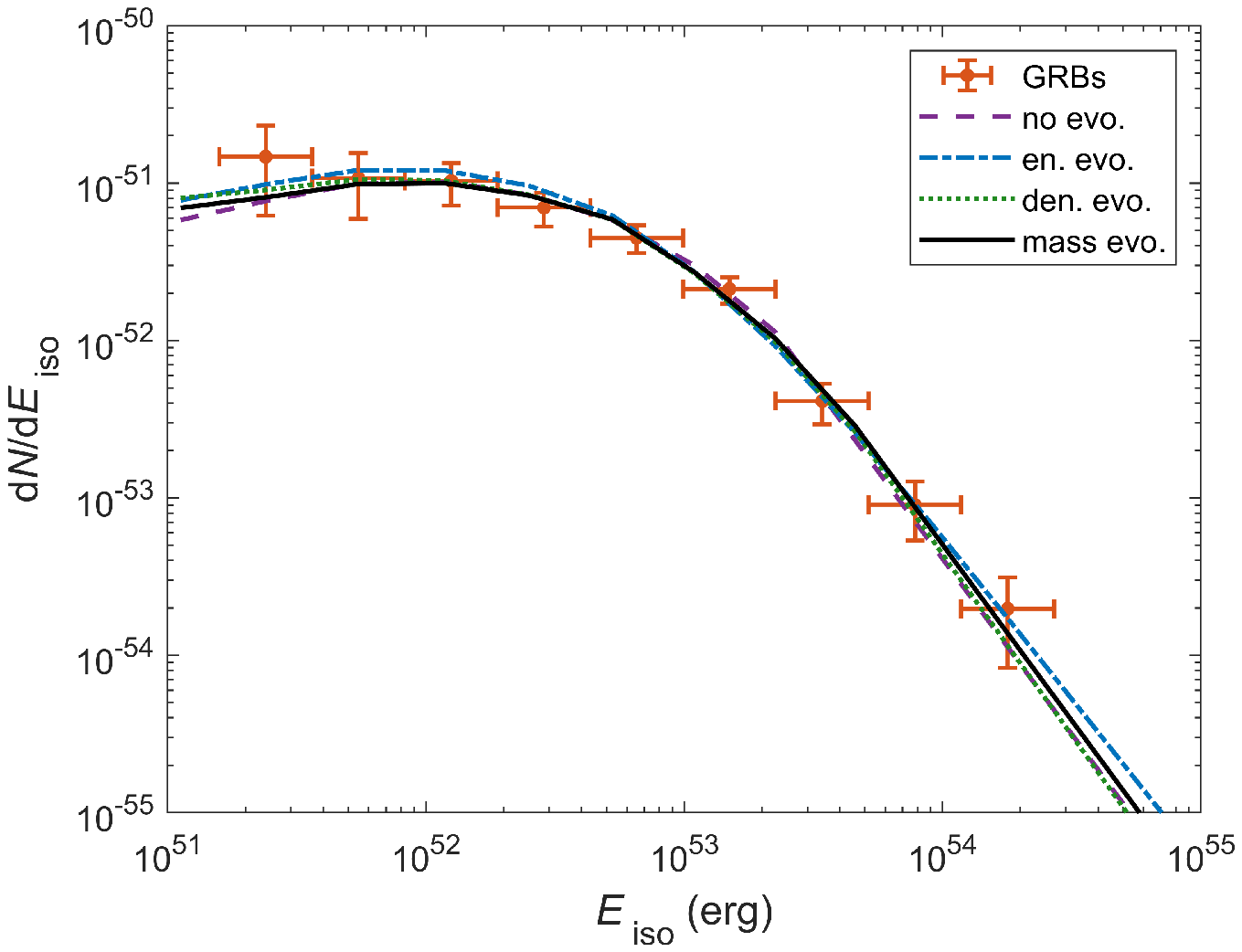}
\includegraphics[angle=0,scale=0.5]{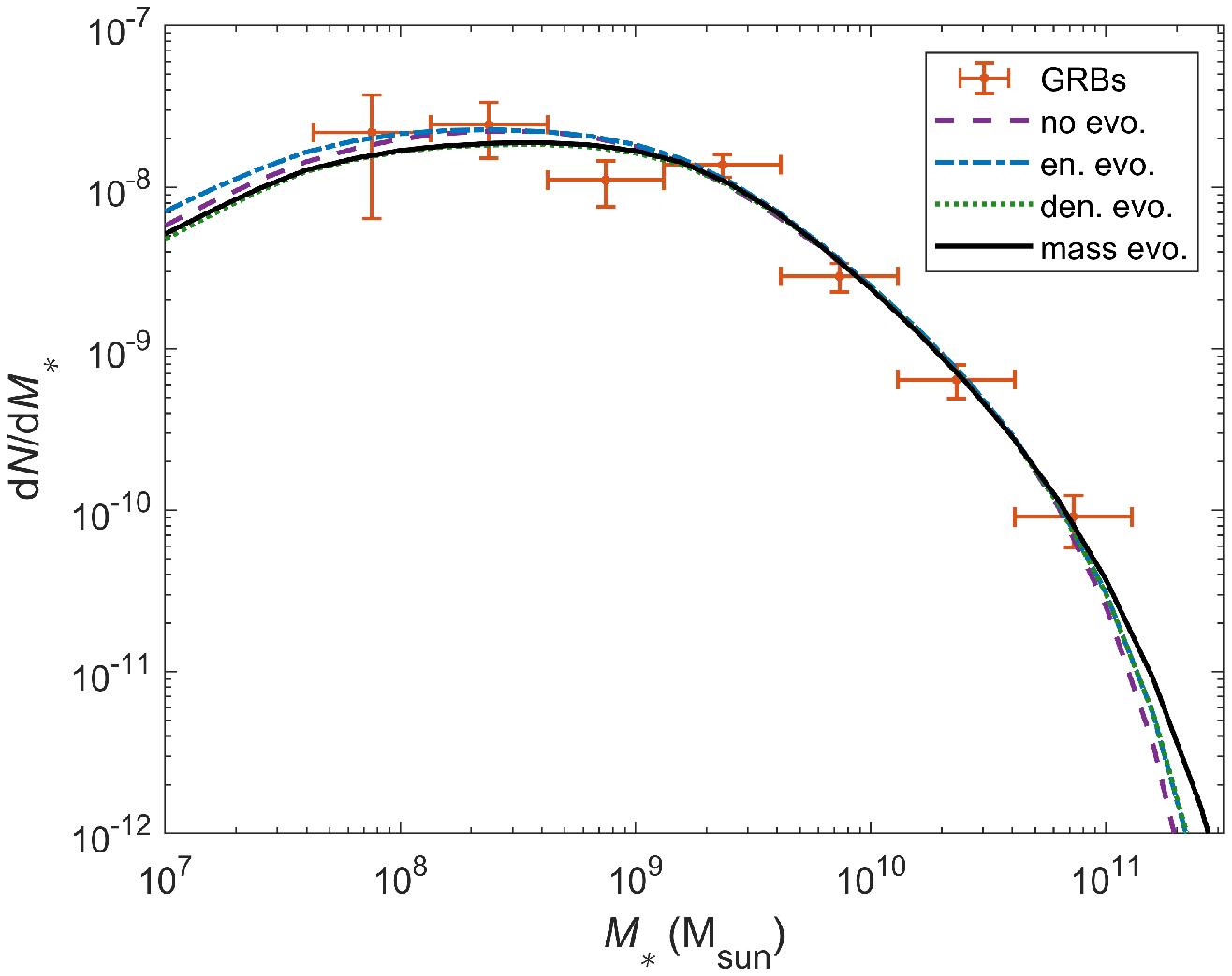}
\caption{Distributions of redshift (top panel), isotropic-equivalent energy (middle panel), and host galaxy mass (bottom panel)
of 109 GRBs in the SHOALS sample (data points, with the number of detections in each corresponding bin indicated by a solid point with Possion error bars). Different curves show the expected distributions from different best-fitting models: no evolution model (purple dashed lines), energy evolution model (blue dotted-dashed lines), density evolution model (green dotted lines), and mass evolution model (black solid lines).}\label{fig2}
\end{figure}

Considering that (i) GRBs may not be good tracers of star formation, (ii) the GRB energy function may evolve
with redshift, and (iii) there is a strong evolution in the galaxy SMF \citep{2006A&A...459..745F,2008ApJ...680...41D,
2014MNRAS.444.2960D}, here we explore four different
evolution scenarios. In the first no evolution model, the GRB formation rate strictly follows the cosmic SFR, and
neither the GRB energy function nor the GRB host SMF evolves with redshift. In the energy evolution model, while
the GRB rate is still proportional to the SFR and the GRB host SMF is still taken to be nonevolving, the break energy
in the GRB energy function increases with redshift as $E_{b,z}=E_{b,0}(1+z)^{\delta}$. In the density evolution model,
the GRB rate follows the SFR in conjunction with an extra evolving factor, i.e., $\psi(z)=\eta \psi_\star(z)(1+z)^{\delta}$,
and both the GRB energy function and the GRB host SMF are taken to be nonevolving.
In the last mass evolution model, while the GRB rate is still fully consistent with the SFR and the break energy
in the GRB energy function is still a constant, the GRB host SMF evolves with redshift as
the following relations \citep{2008ApJ...680...41D}:
\begin{eqnarray}\label{eq:MF-evo}
\varphi(M_{\ast}, z)&=& \varphi_{*}(z)\left(\frac{M_{\ast}}{M_{b}(z)}\right)^{\xi} \exp \left(-\frac{M_{\ast}}{M_{b}(z)}\right) \frac{1}{M_{b}(z)}\nonumber\\
\varphi_{*}(z)&=&\varphi_{0}(1+z)^{\delta}\;,\\
\log M_{b}(z)&=&\log M_{b,0}+\gamma\ln (1+z)\nonumber
\end{eqnarray}
where $\delta$ and $\gamma$ are two additional evolution parameters. When $\delta\rightarrow0$ and $\gamma\rightarrow0$,
Equation~(\ref{eq:MF-evo}) reduces to Equation~(\ref{eq:MF}).

We optimize the model-free parameters by maximizing the likelihood function (Equation~(\ref{eq:Likelihood})).
Since there are several parameters in our models, the Markov Chain Monte Carlo (MCMC) technology is
applied to give multidimensional parameter constraints from the observational data. Here we employ the MCMC code
from GWMCMC,\footnote{https://github.com/grinsted/gwmcmc} which is an implementation of the \citealt{2010CAMCS...5...65G}
Affine invariant ensemble MCMC sampler \citep{2010CAMCS...5...65G,2013PASP..125..306F}.

\section{Model Results}
\label{sec:result}

\newcommand{\tabincell}[2]{\begin{tabular}{@{}#1@{}}#2\end{tabular}}
\begin{table*}
\centering
\caption{Best-fitting parameters in different models with the sample of 109 GRB host galaxies}
\label{tab1}
\resizebox{\textwidth}{!}{
\hspace{-2cm}
\begin{tabular}{lccccccccc}
\hline
  Model &  \tabincell{c}{$\log \eta$} & $a$ & $b$ & $\log E_b$  & $\xi$& $\log M_{b,0}$& Evolution parameters & $\ln\mathcal{L}$ & AIC \\
        &                        \tabincell{c}{(${\rm M}_{\odot}^{-1}$)} &   &  & (erg) & & $({\rm M}_\odot)$& & &  \\
\hline
  No evolution & $-7.82^{+0.38}_{-0.19}$ &$-0.06^{+0.01}_{-0.22}$ & $-1.23^{+0.18}_{-1.45}$& $53.27^{+0.70}_{-0.02}$ & $-0.96^{+0.08}_{-0.08}$ & $10.59^{+0.14}_{-0.07}$ & $\cdot\cdot\cdot$ & -115.17 & 242.34\\
  Energy evolution & $-7.52^{+0.58}_{-0.76}$ & $-0.10^{+0.03}_{-0.47} $ & $-1.07^{+0.39}_{-0.55} $ & $51.91^{+1.68}_{-0.27}$ & $-1.09^{+0.21}_{-0.10} $ & $10.68^{+0.23}_{-0.20}$ & $\delta=2.47^{+0.73}_{-0.89}$ & -106.56 & 227.12\\
  Density evolution & $-7.70^{+0.97}_{-0.29} $& $ -0.46^{+0.34}_{-0.04} $ & $ -1.30^{+0.55}_{-0.94}$ & $53.52^{+0.55}_{-0.67}$ & $-1.10^{+0.06}_{-0.14}$ & $10.69^{+0.23}_{-0.08}$ & $\delta=1.82^{+0.22}_{-0.59}$ & -104.08 & 222.16\\
  Mass evolution & $-7.49^{+0.46}_{-0.59}$ & $-0.39^{+0.18}_{-0.12}$ & $-1.23^{+0.17}_{-1.17}$ & $53.51^{+0.54}_{-0.25}$ & $-1.14^{+0.10}_{-0.06}$ & $11.24^{+0.25}_{-0.43}$ & $\delta=1.96^{+0.44}_{-0.47}$, $\gamma=-0.38^{+0.28}_{-0.19}$ & -103.35 & 222.70\\
  \hline
\end{tabular}}
\begin{description}
  \item[\emph{Note.}] {Errors show the 68\% containment regions around the best-fitting values.}
\end{description}
\label{tab1}
\end{table*}

Using the above analysis method, we can infer the values of each model's free parameters,
including the GRB formation efficiency $\eta$, the GRB energy function, the host galaxy SMF, and
the evolution parameters. The best-fitting parameter values together with their $1\sigma$ uncertainties for different
models are listed in Table~\ref{tab1}. The last two columns report the natural-logarithmic-likelihood value
$\ln\mathcal{L}$ and the
Akaike information criterion (AIC) score, respectively. The AIC score of each fitted model is evaluated as
\citep{1974ITAC...19..716A,2007MNRAS.377L..74L}
\begin{equation}
{\rm AIC}=-2\ln\mathcal{L}+2f\;,
\end{equation}
where $f$ is the number of model-free parameters. If there are two or more models for the same data,
$\mathcal{M}_{\rm 1}, \mathcal{M}_{\rm 2},..., \mathcal{M}_{\rm N}$, and they have been separately
fitted, the one with the smallest AIC score is the one most preferred by this criterion.
With the${\rm AIC}_i$ characterizing model $\mathcal{M}_i$,
the unnormalized confidence that this model is true is the Akaike weight $\exp(-{\rm AIC}_{i}/2)$.
The relative probability of $\mathcal{M}_i$ being the correct model is then
\begin{equation}
P(\mathcal{M}_i)=\frac{\exp(-{\rm AIC}_i/2)}{\exp(-{{\rm AIC_1}}/2)+\cdot\cdot\cdot+\exp(-{\rm AIC_N}/2)}\;.
\end{equation}

\subsection{No evolution model}

In the no evolution scenario, long GRBs are assumed to trace the cosmic star formation and both the GRB energy function and
the GRB host SMF are taken to be nonevolving. Figure~\ref{fig2} shows the observed distributions of redshift $z$,
isotropic-equivalent energy $E_{\rm iso}$, and host galaxy mass $M_{\ast}$ of 109 GRBs in the SHOALS sample. The results
of our fitting from the no evolution model are indicated with purple dashed lines. Although the fits of the $E_{\rm iso}$
and $M_{\ast}$ distributions look good, the expectation from this model does not provide a good representation of the observed
$z$ distribution of the SHOALS sample. In particular, the expected $z$ distribution is found to peak at a lower redshift than
observed, and the rate of GRBs at high-$z$ is underestimated. According to the AIC model selection criterion,
the no evolution model can be safely excluded as having a probability of only $\sim10^{-5}$ of being correct compared to
the other three models (more fully described below).

\subsection{Energy evolution model}
Evolution in the GRB energy function can lead to an enhancement of the high-$z$ GRB detection, providing a
feasible way to reconcile model results with the observations. In this model, we consider the possibility that the break
energy is an increasing function of the reshift, which is $E_{b,z}=E_{b,0}(1+z)^{\delta}$. Note that here the GRB rate
is still proportional to the SFR and the GRB host SMF is assumed to be nonevolving. We find that a strong energy
evolution with $\delta=2.47^{+0.73}_{-0.89}$ is required to reproduce the observed distributions (blue dotted-dashed lines
in Figure~\ref{fig2}). Using the AIC model selection criterion, we find that among the evolution models, this one
is somewhat disfavoured statistically with a relative probability of $\sim5\%$.

\subsection{Density evolution model}
An increase in the GRB formation rate with redshift can also enhance the number of GRB detections at high-$z$.
Here, we assume that the GRB rate follows the SFR in conjunction with an extra evolving factor characterized by
$(1+z)^{\delta}$, i.e., $\psi(z)=\eta \psi_\star(z)(1+z)^{\delta}$. Note that in this model both the GRB energy function
and the GRB host SMF are taken to be nonevolving. We find that a strong density evolution with $\delta=1.82^{+0.22}_{-0.59}$
reproduces the observed distributions (green dotted lines in Figure~\ref{fig2}) quite well.
The AIC shows that the density evolution model is statistically preferred with a relative probability of $\sim54\%$.

\subsection{Mass evolution model}
Here we investigate whether an evolving SMF of GRB hosts can reconcile the discrepancy between the GRB rate and the SFR.
In this model, while the GRB formation rate still follows the cosmic SFR and the GRB energy function also does
not evolve with redshift, the Schechter parameters of the GRB host SMF are assumed to be evolving with redshift
as $\varphi_{*}(z)=\varphi_{0}(1+z)^{\delta}$ and $\log M_{b}(z)=\log M_{b,0}+\gamma\ln (1+z)$. The fitting results
from the mass evolution model are shown as the black solid lines in Figure~\ref{fig2}, which agree with the
observed distributions very well. That is, a strong redshift evolution in the GRB host SMF (with evolution indices of
$\delta=1.96^{+0.44}_{-0.47}$ and $\gamma=-0.38^{+0.28}_{-0.19}$) is required to reproduce the observations well.
We find that $\delta=0$ is ruled out at $\sim4.2\sigma$ level and $\gamma=0$ is ruled out at $\sim1.4\sigma$ level,
confirming the significant dependence of Schechter parameters on redshift. Using the galaxy stellar mass data
spanning $0<z<5$ from the FORS Deep Field, \cite{2008ApJ...680...41D} also found a strong evolution in mass,
in good agreement with our results from the GRB host galaxies. According to the AIC, the mass evolution model
is slightly disfavoured compared to the density evolution model, but the differences are statistically insignificant
($\sim41\%$ for the former versus $\sim54\%$ for the latter).

It should be underlined that the $(1+z)^{\delta}$ term in the GRB host SMF (see Equation~(\ref{eq:MF-evo}))
can be referred to as the GRB formation efficiency $\eta$ instead of the normalization $\varphi_{*}(z)$. In other words, this
term can be taken out from the mass integral in Equation~(\ref{eq:MF-evo}) as it does not depend on the stellar mass.
However, in addition to this evolutionary term, a modest redshift evolution in the break mass of the SMF (with an evolution
index of $\gamma=-0.38^{+0.28}_{-0.19}$) is identified by the observations, serving as an important characteristic for
distinguishing between the density evolution model and the mass evolution model.

\begin{table*}
\centering
\caption{Best-fitting parameters in Different Models Using the Refined Sample of 109 GRB hosts (Including 79 hosts with known stellar masses, 26 with mock stellar masses, and 4 with stellar-mass upper limits)}
\label{tab2}
\resizebox{\textwidth}{!}{
\hspace{-2cm}
\begin{tabular}{lccccccccc}
\hline
  Model &  \tabincell{c}{$\log \eta$} & $a$ & $b$ & $\log E_b$  & $\xi$& $\log M_{b,0}$& Evolution parameters & $\ln\mathcal{L}$ & AIC \\
        &                        \tabincell{c}{(${\rm M}_{\odot}^{-1}$)} &   &  & (erg) & & $({\rm M}_\odot)$& & &  \\
\hline
  No evolution& $-7.63^{+0.48}_{-0.27} $ &$-0.14^{+0.11}_{-0.11} $ & $-1.38^{+0.29}_{-0.62}$& $53.44^{+0.24}_{-0.20}$ & $-1.02^{+0.09}_{-0.08}$ & $10.67^{+0.13}_{-0.12}$ & $\cdot\cdot\cdot$ & -114.46 & 240.92\\
  Energy evolution & $-7.30^{+1.01}_{-0.44}$ & $-0.36^{+0.30}_{-0.20}$ & $-1.06^{+0.29}_{-0.30} $ & $52.12^{+1.03}_{-0.62}$ & $-1.16^{+0.12}_{-0.12}$ & $10.78^{+0.30}_{-0.14}$ & $\delta=2.72^{+0.80}_{-0.63}$ & -105.06 & 224.12\\
  Density evolution & $-7.00^{+0.88}_{-0.64}$ & $-0.45^{+0.15}_{-0.10} $ & $-2.26^{+0.99}_{-1.31}$ & $54.03^{+0.22}_{-0.56}$ & $-1.22^{+0.12}_{-0.11}$ & $10.79^{+0.27}_{-0.16}$ & $\delta=1.71^{+0.48}_{-0.33}$ & -102.15 & 218.30\\
  Mass evolution & $-6.86^{+0.57}_{-0.92}$ & $-0.24^{+0.14}_{-0.15} $ & $-1.47^{+0.46}_{-0.71} $ & $53.48^{+0.49}_{-0.32}$ & $-1.22^{+0.11}_{-0.12}$ & $11.38^{+0.43}_{-0.49}$ & $\delta=2.03^{+0.67}_{-0.67}$, $\gamma=-0.45^{+0.36}_{-0.37}$ & -100.63 & 217.26\\
  \hline
\end{tabular}}
\end{table*}

\begin{figure}
\center
\includegraphics[angle=0,scale=0.5]{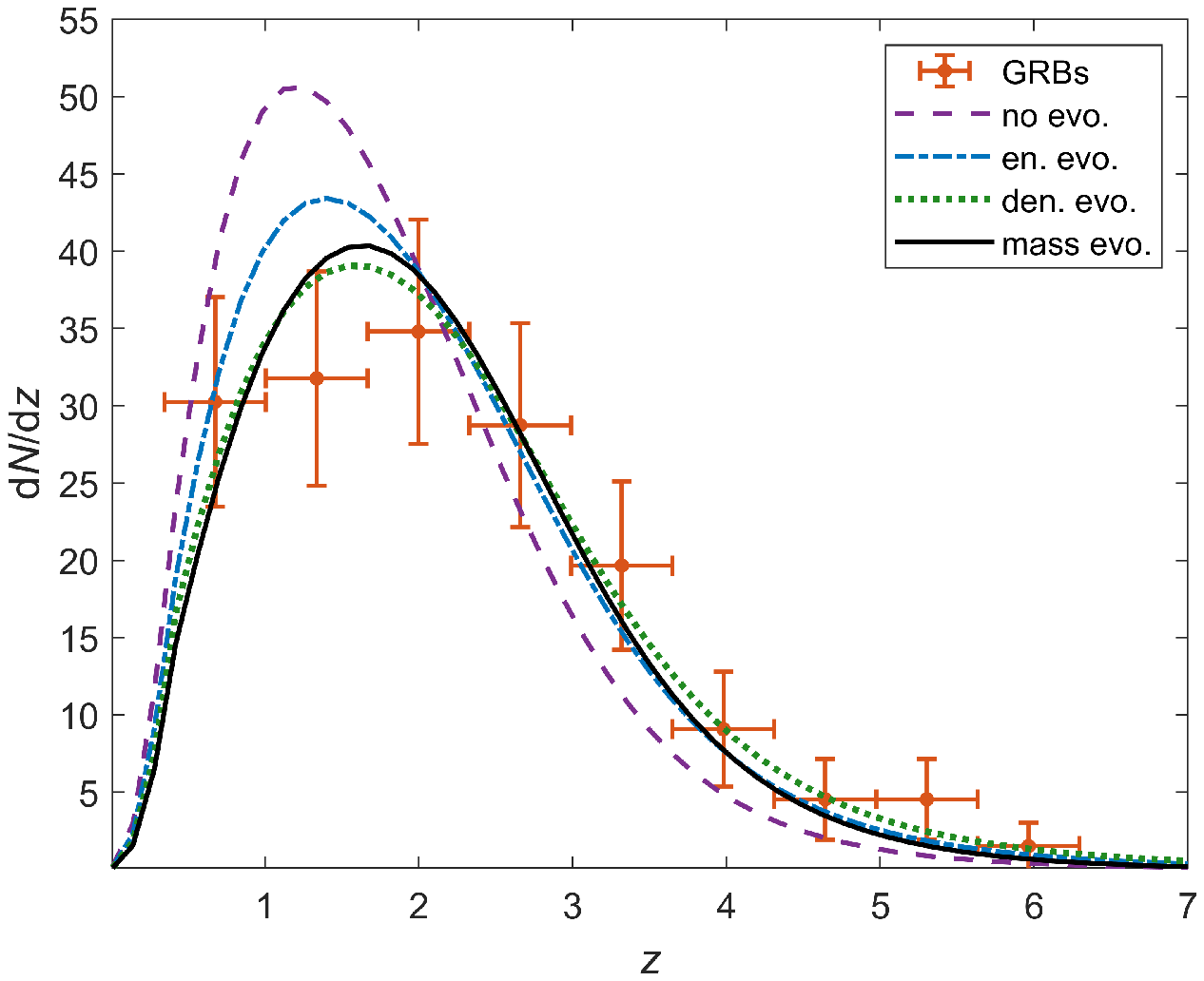} 
\includegraphics[angle=0,scale=0.5]{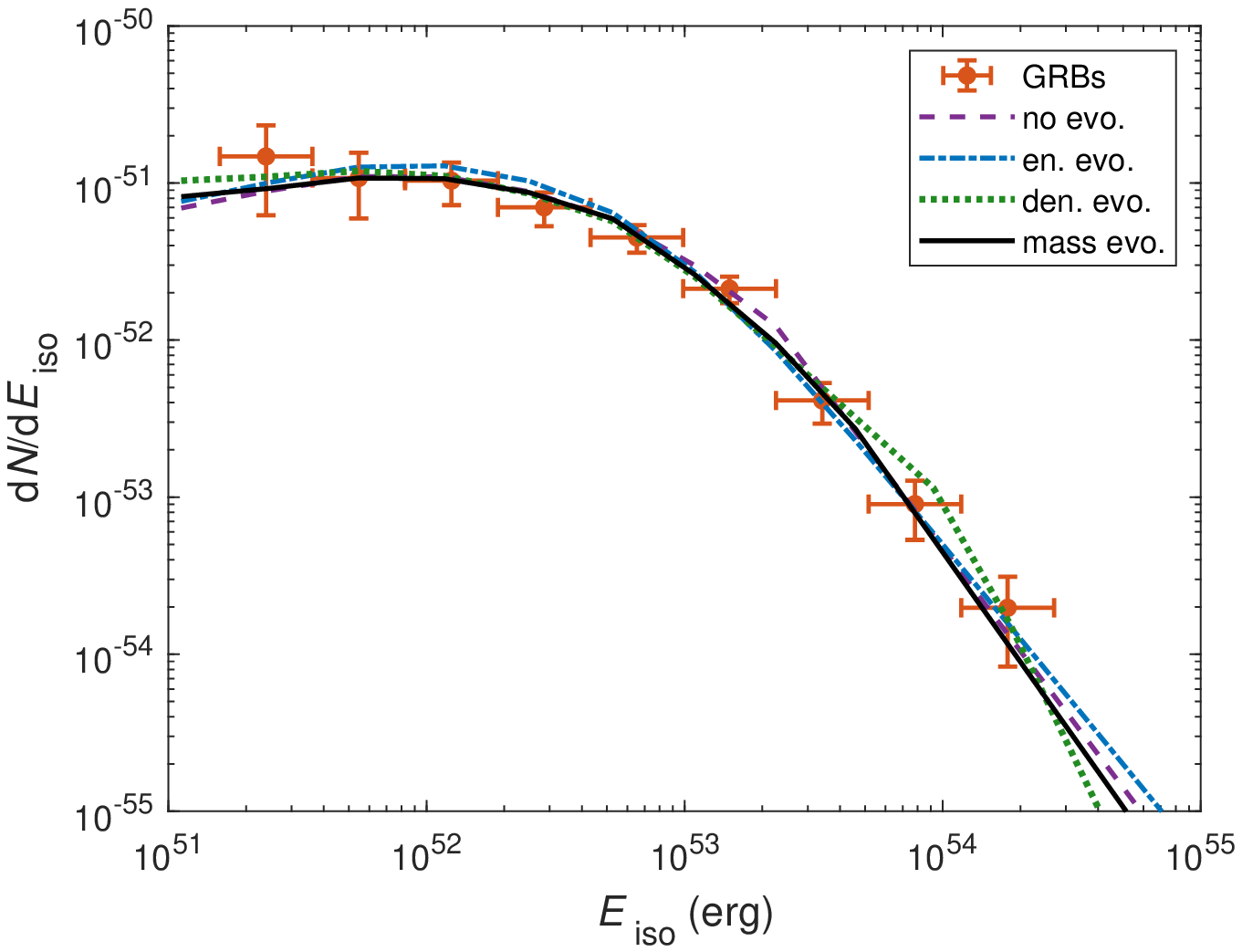} 
\includegraphics[angle=0,scale=0.5]{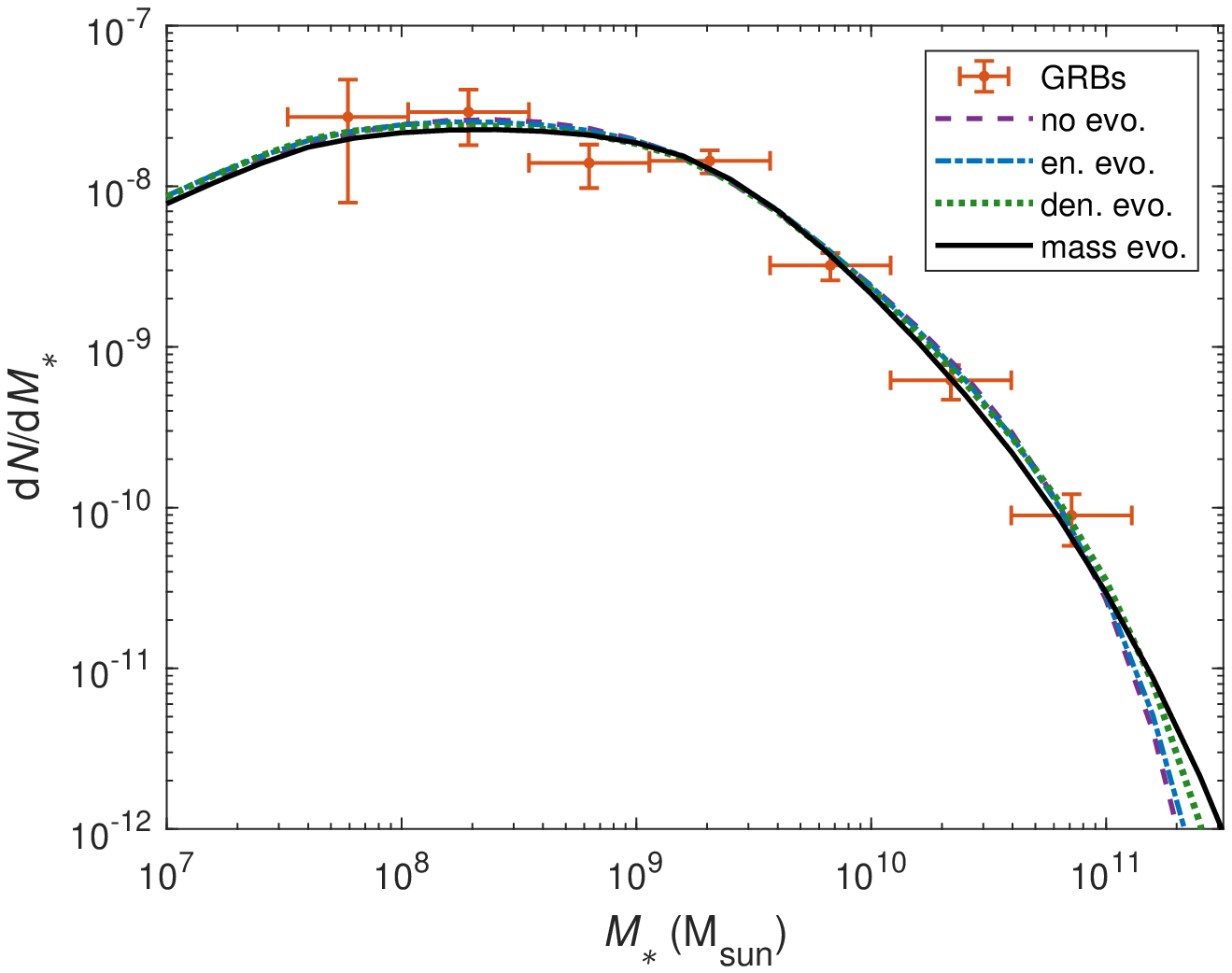} 
\caption{Same as Figure~\ref{fig2}, but for the refined sample of 109 GRB hosts (including 79 hosts with known stellar masses, 26 with mock stellar masses, and 4 with stellar-mass upper limits). Model results are shown as in Table~\ref{tab2}.}\label{fig3}
\end{figure}

Note that there are 30 GRB hosts with stellar-mass upper limits in our analysis sample, and we simply treated these
upper mass limits as the stellar-mass measurements. To examine whether these upper mass limits can reliably be included for our analysis, we perform Monte Carlo simulations on their stellar masses. For each GRB host with measured $z$ and stellar-mass upper limit $M_{\ast,\rm upper}$, we mock the stellar mass $M_{\ast,\rm mock}$ according to the mass probability distribution of 79 hosts having measured stellar masses. With the measured $z$, we infer the stellar-mass threshold $M_{\ast,\rm lim}$ using Equations~(\ref{eq:MAB}) and (\ref{eq:M-MAB}). If $M_{\ast,\rm lim}(z)<M_{\ast,\rm mock}<M_{\ast,\rm upper}$, we pick up this simulated event as a triggered GRB host with mass detection. Since there
are four GRB hosts (GRBs 060522, 060927, 081029, and 091208B) with $M_{\ast,\rm upper}\leq M_{\ast,\rm lim}(z)$, their upper mass limits are still treated as the stellar-mass measurements. While the stellar masses of the remaining 26 are generated through Monte Carlo simulations, the simulation process is repeated 1000 times to ensure the final constraints are unbiased.

Following the same analysis method described in Section~\ref{sec:method}, we also use the refined sample of 109 GRB hosts (including 79 hosts with measured $M_{\ast}$, 26 with mock $M_{\ast,\rm mock}$, and 4 with $M_{\ast,\rm upper}$) to construct the host galaxy SMF in different evolution models. The best-fitting results are displayed in Table~\ref{tab2} and Figure~\ref{fig3}. We find that the expectations from our evolution models give a good description of the observed distributions of the refined sample, whereas the no evolution model provides a poor match for the observed redshift distribution. According to the AIC, we can safely discard the no evolution model as having
a probability of $\sim10^{-6}$ of being correct compared to the other three models. This further verifies the reliability of our analysis method, strengthening our conclusions. That is, the inclusion of 30 GRB hosts with stellar-mass upper limits would not change the main conclusions of our paper.

\section{Summary}
\label{sec:summary}

Whether long GRBs are biased tracers of the star formation activity remains an open question. Studies of
the GRB redshift distribution seem to indicate that the comoving GRB rate density exhibits similar behavior to
the star formation history at low redshifts ($z\la2$), but shows an enhancement of GRBs at high redshifts
($z\ga2-3$) compared to what would be expected if the GRB rate followed the cosmic SFR strictly (e.g.,
\citealt{2006ApJ...647..773D,2008ApJ...673L.119K,2015ApJ...801..102P}). However, in the technique of star formation models,
the GRB rate measured at any epoch is an average over all galaxies at that time (a diverse galaxy population
spanning orders of magnitude in metallicities, SFRs, etc.). Thus, it is difficult to inspect which types
of galaxies contribute most to the GRB formation rate. To get a better understanding of the connection
between GRB production and environment, one has to characterize the population of GRB host galaxies
directly \citep{2016ApJ...817....8P}.

Based on the complete GRB host sample presented in \cite{2016ApJ...817....7P,2016ApJ...817....8P}, we construct
the SMF of GRB host galaxies as well as the isotropic-equivalent energy function and redshift distribution of
GRBs in the frameworks of different evolution models. The maximum likelihood method is applied to perform
a joint analysis for the distributions of redshift, isotropic-equivalent energy, and host galaxy mass of
the GRB sample, and the MCMC algorithm is adopted to provide the best-fitting parameters in each model.
Our results confirm that GRBs must have experienced some kind of redshift evolution, being more luminous
or more population in the past than present day. In order to match the observed distributions, the typical GRB
energy should increase to $(1+z)^{2.47^{+0.73}_{-0.89}}$ or the GRB rate density should rise to $(1+z)^{1.82^{+0.22}_{-0.59}}$
on top of the known cosmic evolution of the SFR. We find that the GRB host SMF can be well described by the
Schechter function with a power-law index $\xi\approx-1.10$ and a break mass $M_{b,0}\approx4.9\times10^{10}$ ${\rm M}_\odot$,
independent of the assumed evolutionary effects.

However, since the SMF of all galaxies is suggested to be evolving with redshift, the intrinsic form of the stellar-mass distribution of GRB host galaxies may also have a redshift dependence. We also explore models in which the GRB host SMF
evolves with redshift. In this case, we find that the redshift distribution of GRBs with measured redshift, energy,
and stellar mass can be successfully fitted by the star formation history, which involves with an evolving SMF of GRB hosts.
In order to reproduce the observed redshift distribution well, the normalization coefficient of the SMF should increase to
$(1+z)^{1.96^{+0.44}_{-0.47}}$ and the break mass should evolve with redshift according to
$\log M_{b}(z)=11.24^{+0.25}_{-0.43}- 0.38^{+0.28}_{-0.19}\ln (1+z)$. Our results suggest that long GRBs
may be biased tracers of star formation and the evolving SMF may also explain the discrepancy between the GRB rate
and the SFR.

The current sample size of GRB hosts is still too small to exhibit an obvious redshift evolution in the stellar
mass distribution. Ongoing large-area surveys will be able to increase the host galaxy sample size. With more
abundant observational information in the future, the properties of the population of GRB host galaxies and their
possible evolution with redshift will be better characterized, and determinations on the SMF of GRB hosts, as discussed
in this work, will be considerably improved.

\acknowledgments
We thank the anonymous referee for providing constructive comments that have led to
a significant improvement in the presentation of the material in this paper.
This work is partially supported by the National Natural Science Foundation of China
(grant Nos.~11725314, 12041306, and 12103089), the Key Research Program of Frontier Sciences
(grant No. ZDBS-LY-7014) of Chinese Academy of Sciences, the Major Science and Technology
Project of Qinghai Province (2019-ZJ-A10), the China Manned Space Project (CMS-CSST-2021-B11),
the Natural Science Foundation of Jiangsu Province (grant Nos. BK20221562 and BK20211000),
and the Guangxi Key Laboratory for Relativistic Astrophysics.

\bibliographystyle{apj}
\bibliography{apj}

\end{document}